\def\fr#1#2{\hbox{${#1\over #2}$}}

\def\ni{\noindent}
\def\vs{\vskip.3cm}

\def\+{{(+)}}  \def\-{ {(-)} }   \def\0{ {(0)} }
\def\1{ {(1)} }  \def\2{ {(2)} }

                            \def\k{\kappa}

\def\sq{Q\kern-6pt/}
\def\sQ{Q\kern-12pt\nearrow}
\documentclass[11pt,eqs]{article}

\usepackage{latexsym}
\def\be{\begin{equation}}             \def\ee{\end{equation}}
\def\ba{\begin{array}{rcl}}           \def\ea{\end{array}}
\def\beqa{\begin{eqnarray} }          \def\eeqa{\end{eqnarray} }
\def\beqalign{\begin{eqalign}}        \def\eeqalign{\end{eqalign}}
             
\def\bsubeq{\begin{subequations}}     \def\esubeq{\end{subequations}}
\def\bitem{\begin{itemize}}           \def\eitem{\end{itemize}}

\def\DJ{\leavevmode\setbox0=\hbox{D}\kern0pt
 \rlap{\kern.04em\raise.188\ht0\hbox{-}}D}
\def\dj{\leavevmode\setbox0=\hbox{d}\kern0pt
 \rlap{\kern.215em\raise.46\ht0\hbox{-}}d}

\newcommand{\bd}{\begin{displaymath}}
\newcommand{\ed}{\end{displaymath}}
\begin{document}

\title{ Gauge symmetries decrease the number of Dp-brane dimensions
\thanks{Work supported in part by the Serbian Ministry of Science and
Environmental Protection, under contract No. 141036.}}
\author{B. Nikoli\'c \thanks{e-mail address: bnikolic@phy.bg.ac.yu} and B. Sazdovi\'c
\thanks{e-mail address: sazdovic@phy.bg.ac.yu}\\
       {\it Institute of Physics, 11001 Belgrade, P.O.Box 57, Serbia}}
%\date{}
\maketitle
\begin{abstract}

It is known that the presence of antisymmetric background field
$B_{\mu\nu}$ leads to noncommutativity of Dp-brane manifold. Addition of the linear dilaton field in the form $\Phi(x)=\Phi_0+a_\mu x^\mu$,
causes the appearance of the commutative Dp-brane coordinate
$x=a_\mu x^\mu$. In the present article we show that for some particular choices of the background fields, $a^2\equiv G^{\mu\nu}a_\mu a_\nu=0$ and $\tilde a^2\equiv \ [ (G-4BG^{-1}B)^{-1}\ ]^{\mu\nu}a_\mu a_\nu=0$, the local
gauge symmetries appear in the theory. They turn some Neuman boundary conditions into the Dirichlet ones, and consequently decrease the number of the Dp-brane dimensions.

\end{abstract}
\vs

\ni {\it PACS number(s)\/}: 11.10.Nx, 04.20.Fy, 11.25.-w   \par
\section{Introduction}

Quantization of the open string ending on the Dp-brane, with constant metric
$G_{\mu \nu}$ and antisymmetric tensor Kalb-Ramond field $B_{\mu \nu}$, leads to the
noncommutativity of the Dp-brane manifold \cite{CDS}.
Inclusion of the dilaton field $\Phi$ linear in $x^\mu$ was studied in
\cite{LRR}-\cite{BS2}. In ref.\cite{LRR}, the Dirichlet boundary conditions were
constructed, while in refs.\cite{BKM,BS2}, the noncommutativity structure was
analyzed. Ref.\cite{BS2} considers the conformal part of the world-sheet
metric $F$ as a dynamical variable, introducing an additional boundary
condition corresponding to $F$. It becomes a new noncommutative variable,
while Dp-brane coordinate in the direction of the dilaton gradient
$a_i=\partial_i \Phi$, turns into a commutative one.

Let us clarify the role of the field $F$ in the open and closed string theory. In presence of the dilaton field, the classical action as well as the Neumann boundary
conditions explicitly depend on $F$ and break conformal invariance. On the space-time field equations the field $F$ decouples on the quantum level from the closed string theory. These conditions do not guarantee that it decouples from the open string theory, because the contribution of the boundary conditions should be investigated. The
purpose of this article is to check whether open string theory depends on the field $F$.

We consider some particular cases when $a_i$ is light-like vector, with respect to the closed string
metric $G_{ij}$, $a^2=G^{ij}a_i a_j=0$, as well as to the effective one $G^{eff}_{ij}$,
$\tilde a^2=(G^{-1}_{eff})^{ij}a_i a_j=0$. As a consequence of these
conditions, the first class constraints and local gauge symmetries appear. These gauge symmetries have different origins. In the first case, $a^2$
is a coefficient in front of $\dot F$, so that condition $a^2=0$ produces the
standard canonical constraint, which is of the first class. In the
second case, some of the constraints originated from the boundary conditions, for $\tilde a^2=0$ turn from the second class into the first
class constraints.

As a consequence of gauge symmetries, some of the initial Dp-brane
coordinates change the
corresponding boundary conditions from Neuman to Dirichlet and
decrease the number of the Dp-brane dimensions. The closed string coordinates, which depend on the open string ones but also on the
corresponding momenta, define the noncommutative subspace of Dp-brane. The noncommutativity
parameter is proportional to the antisymmetric field $B_{ij}$. Some closed string coordinates depend only on the open string ones and they are the
commutative directions of Dp-brane.

In all cases, solving the second class constraints, we find the effective theory.
It is again the string theory defined in terms of the open string variables, symmetric
under transformation $\sigma \to -\sigma$ and propagating in new open string background.

At the end we shortly discuss the another possibility to cancel conformal anomaly by introducing Liouville action, which is kinetic term for the field $F$. The results obtained in this framework are in full correspondence with the results in absence of the Liouville action, which will be discussed in concluding remarks.

The appendix is devoted to some geometrical tools, which help us to express the results clearly.

\section{Action and space-time field equations}
\setcounter{equation}{0}

Let us introduce action describing the open string propagation
in the background defined by the space-time metric
$G_{\mu\nu}(x)$, Kalb-Ramond antisymmetric field $B_{\mu\nu}(x)$
and dilaton scalar field $\Phi(x)$ \cite{P}
\begin{equation}\label{eq:action2}
S= \kappa  \int_\Sigma  d^2 \xi  \sqrt{-g}  \left\{  \left[  {1
\over 2}g^{\alpha\beta}G_{\mu\nu}(x) +{\varepsilon^{\alpha\beta}
\over \sqrt{-g}}  B_{\mu\nu}(x) \right] \partial_\alpha x^\mu
\partial_\beta x^\nu +  \Phi (x) R^{(2)}  \right\} \,  .
\end{equation}
Here, $x^{\mu}$ ($\mu=0,1,2,...,D-1$) are coordinates of the $D$ dimensional
space-time and $\xi^{\alpha}(\alpha=0,1)$ parameterize the two dimensional
world-sheet $\Sigma$. Further, $R^{(2)}$ denotes the scalar curvature related
to the intrinsic world-sheet metric $g_{\alpha\beta}$, while $g=\det g_{\alpha\beta}$. The antisymmetric
tensor $\varepsilon^{\alpha\beta}$ is two dimensional $\varepsilon$-symbol
with the adopted convention $\varepsilon^{01}=-1$. The constant $\kappa=\frac{1}{2\pi\alpha'}$
is known as string tension, where $\alpha'$ is Regge slope. The following notation is
used: $\partial_{\alpha}=\frac{\partial}{\partial\xi^{\alpha}}$,
$\partial_{\mu}=\frac{\partial}{\partial x^{\mu}}$, $\xi^{0}=\tau$ and
$\xi^{1}=\sigma$   ($\sigma\in\left[ 0,\pi\right]$).

The goal of introducing dilaton field is to preserve classical Weyl
symmetry on the quantum level \cite{FT}. It is realized under the following
conditions \cite{FCB}
\begin{equation}\label{eq:betaG}
\beta^G_{\mu \nu} \equiv  R_{\mu \nu} - \fr{1}{4} B_{\mu \rho
\sigma}
B_{\nu}{}^{\rho \sigma} +2 D_\mu a_\nu =0   \,  ,
\end{equation}
\begin{equation}\label{eq:betaB}
\beta^B_{\mu \nu} \equiv  D_\rho B^\rho{}_{\mu \nu} -2 a_\rho
 B^\rho{}_{\mu \nu} = 0   \,  ,
\end{equation}
\begin{equation}\label{eq:betaFi}
\beta^\Phi \equiv 4 \pi \kappa {D-26 \over 3} -
 R + \fr{1}{12} B_{\mu \rho \sigma}
B^{\mu \rho \sigma} - 4 D_\mu a^\mu + 4 a^2 = 0  \,  ,
\end{equation}
where $R_{\mu \nu}$, $R$ and $D_\mu$ are space-time Ricci tensor, scalar
curvature and covariant derivative, respectively, $B_{\mu \rho \sigma}$ is
field strength of the field $B_{\mu \nu}$ and the vector $a_\mu$ is the
gradient of the dilaton field.

The background fields in the form \cite{P}
$$
G_{\mu \nu}(x) = G_{\mu \nu}=const \,  , \quad B_{\mu \nu} (x) =
B_{\mu \nu} = const\, ,
$$
$$
\Phi(x) =\Phi_0 + a_\mu x^\mu  \,  , \quad (a_\mu = const)
$$
are an exact solution of the above equations for
\begin{equation}\label{eq:dimenzije}
a^2 = \kappa \pi {26 - D \over 3}  \,  .
\end{equation}

Note that for $a^2 =0$ the theory describes the critical string ($D=26$) and for
$a^2 > 0$ the noncritical one ($ D < 26$).

By $x^i(\xi) \,  (i =0,1,...,p)$ we denote Dp-brane
coordinates and by $x^a(\xi) \,  (a = p+1,p+2,...,D-1)$ the orthogonal ones so
that $G_{\mu \nu} = 0$ for $\mu =i$ and $\nu = a$. The part of the action (\ref{eq:action2}) describing the
free string propagation in $x^a$ directions decouples from the rest. In order to simplify the problem, we assume that the background fields $B_{\mu
\nu}$ and $a_\mu$ are nontrivial only along Dp-brane: $B_{\mu \nu} \to B_{i
j}$, $a_\mu \to a_i$.

Choosing the conformal gauge $g_{\alpha \beta} = e^{2 F} \eta_{\alpha
\beta}$, we obtain
\begin{equation}\label{eq:2deo}
S(x^i,F) =\k \int_{\Sigma}d^2 \xi\bigg [\bigg
(\frac{1}{2}\eta^{\alpha\beta}G_{ij}+\epsilon^{\alpha\beta} B_{ij}
\bigg)\partial_{\alpha}x^{i}\partial_{\beta}x^{j}+2\eta^{\alpha\beta}a_i
\partial_{\alpha}x^i\partial_{\beta}F\bigg] \,  .
\end{equation}
Classically, it is not conformally invariant, because it depends on the
conformal part of the metric, $F$.

\section{Noncommutativity in the case $a^2\neq 0$ and $\tilde a^2\neq0$}
\setcounter{equation}{0}

In order to introduce notation and to have possibility to discuss all cases from an unique point
of view, let us make a brief review of refs.\cite{BS2,BSijmp}.

The canonical Hamiltonian
\begin{equation}\label{eq:hamiltonijan}
H_{c}=\int_{0}^{\pi}d\sigma\mathcal{H}_{c}= \int_0^{\pi} d\sigma (T_{-}-T_{+})  \,  ,
\end{equation}
is expressed in terms of the energy-momentum tensor components
\begin{equation}\label{eq:enmomten}
T_{\pm}=\mp\frac{1}{4\kappa}\left( G^{ij}J_{\pm i}J_{\pm
j}+\frac{j}{a^2}i^{\Phi}_{\pm}\right)+\frac{1}{2}\left(
{i^\Phi_\pm}'-F'i^\Phi_\pm \right) \,  .
\end{equation}

The currents on the Dp-brane are
\begin{equation}\label{eq:Jstruja}
J^i_{\pm}=(P_T^0)^{ij}j_{\pm j}+\frac{a^i}{2 a^2}i^\Phi_{\pm}=j^i_\pm-\frac{a^i}{a^2}j\, ,
\end{equation}
\begin{equation}\label{eq:jistruja}
j_{\pm i}=\pi_i+2\kappa \Pi_{\pm ij}x'^j\, ,\quad \bigg (\Pi_{\pm ij}=
B_{ij}\pm \frac{1}{2}G_{ij}\bigg )
\end{equation}
\begin{equation}\label{eq:jstruja}
i^\Phi_\pm= \pi \pm 2\kappa a_i x'^i\, ,  \qquad j= a^i j_{\pm
i}-\frac{1}{2}i^\Phi_{\pm}  \, ,
\end{equation}
\begin{equation}\label{eq:Fstruja}
i^F_{\pm}=\frac{a^i}{a^2}j_{\pm i}-\frac{1}{2a^2}i^\Phi_{\pm}\pm 2\kappa F'\, ,
\end{equation}
where the projector $(P_T^0)_i{}^j$ is defined in (\ref{eq:ptpl}). We use the notation $\partial_{\sigma} X= X'$, for any variable $X$. The canonical momenta conjugated to $x^i$ and $F$ are denoted by $\pi_i$ and $\pi$, respectively. The Poisson brackets between canonical Hamiltonian and the currents $J_{\pm i}$, $i^\Phi_{\pm}$ and $i^F_\pm$ are
\begin{equation}\label{eq:hamstruja}
\left\lbrace  H_c, J_{\pm i}\right\rbrace =\mp J'_{\pm i}\, ,\quad \left\lbrace  H_c, i^\Phi_{\pm}\right\rbrace =\mp i'^\Phi_{\pm}\, ,\quad \left\lbrace  H_c, i^F_{\pm}\right\rbrace =\mp i'^F_{\pm}\, .
\end{equation}

\subsection{Boundary conditions and consistency procedure}

The Neuman boundary conditions corresponding to the fields $x^i$ and $F$ are
of the form $\gamma_{i}^{(0)} \Big |_0^\pi=0$ and $\gamma^{(0)}\Big
|_0^\pi=0$, where
\begin{equation}\label{eq:gruslov}
\gamma_{i}^{(0)}=\Pi_{+ ij}J_{-}^j+\Pi_{-ij}J^{j}_{+}+\frac{a_i}{2}\left(i^F_{-}-i^F_{+} \right)   \,  ,
\quad \gamma^{(0)}=\frac{1}{2}\left( i^\Phi_{-}-i^\Phi_{+} \right)\, .
\end{equation}
They are considered as canonical constraints. Assuming that $G_{i j}$,
$B_{i j}$ and $a_i$ are constant and applying the standard Dirac procedure, with the help of the relations (\ref{eq:hamstruja}), we obtain a set of compact conditions

\begin{equation}\label{eq:velikog}
\Gamma_{i}(\sigma)=\Pi_{+ ij}J_{-}^j(\sigma)+\Pi_{-ij}J^{j}_{+}(-\sigma)+
\frac{a_i}{2}\left[i^F_{-}(\sigma)-i^F_{+}(-\sigma) \right] \,  ,
\end{equation}
\begin{equation}\label{eq:malog}
\Gamma(\sigma)=\frac{1}{2}\left[i^\Phi_-(\sigma)-i^\Phi_+(-\sigma)\right]\, ,
\end{equation}
and conclude that the canonical variables are $2\pi$ periodic functions.

The Poisson brackets between canonical Hamiltonian and constraints $\Gamma_i$
and $\Gamma$ are equal to their sigma derivatives, which means that there are no more constraints in the theory and we completed the consistency procedure.

The complete set of the constraint algebra is of the form
\begin{equation}\label{eq:dklasa}
\left\lbrace \chi_A(\sigma), \chi_B(\overline \sigma) \right\rbrace=-\kappa M_{AB}\delta'\, ,\quad \chi_A=(\Gamma_i,\Gamma)\, , \quad \Big[ \delta'=\partial_\sigma \delta(\sigma-\overline \sigma)\Big] \, , 
\end{equation}
where
\begin{equation}\label{eq:matM}
M_{AB}=\left (
\begin{array}{rr}
\tilde G_{ij} & 2a_i\\
2a_j & 0
\end{array}\right )\, .
\end{equation}
To the space-time component
\begin{equation}\label{eq:effm}
\tilde G_{ij}=G_{ij}-4B_{ik}(P_T^0)^{kl}B_{lj}=(\hat P_T^1 G_{eff})_{ij}\, , \quad
(G^{eff}_{ij}=G_{ij}-4B_{ik}G^{kl}B_{lj})\, ,
\end{equation}
we will refer as the open string metric, where $(\hat P_T^1)_i{}^j$ is defined in eq.(\ref{eq:kapapet1}).

The determinant
\begin{equation}\label{eq:detM}
\det M_{AB}=-4\tilde a^2 \det \tilde G_{ij}\, ,
\end{equation}
shows that for $\tilde a^2\neq0$ and $a^2\neq0$, which we assume in this section, the rank of
$M_{AB}$ is equal to $p+2$ and all constraints are of the second class (except
the zero modes, see \cite{BS3}).

\subsection{Solution of the second class constraints and noncommutativity}

In terms of the open string variables
\begin{equation}\label{eq:mena1}
q^i(\sigma)=\frac{1}{2}\left[x^i (\sigma)+x^i (-\sigma)\right] \,  ,
\quad \overline{q}^{i}(\sigma)=\frac{1}{2}\left[
x^{i}(\sigma)-x^{i}(-\sigma)\right] \, ,
\end{equation}
\begin{equation}\label{eq:mena2}
f(\sigma)=\frac{1}{2}\left[
F(\sigma)+F(-\sigma)\right] \,  , \quad
\overline{f}(\sigma)=\frac{1}{2}\left[
F(\sigma)-F(-\sigma)\right] \,  ,
\end{equation}
(with similar expressions for momenta $p_i$, $\overline p_i$, $p$ and $\overline p$),
the constraints $\Gamma_{i}(\sigma)$
and $\Gamma(\sigma)$ take the form
\begin{equation}\label{eq:veza1}
\Gamma_i=2{(BP_T^0)_i}^j p_j+\frac{1}{a^2}(Ba)_i p-\k
\tilde G_{ij}{\overline q'}^j-2\k a_i {\overline f}'+\overline p_i\, ,\quad \Gamma=
\overline p-2\kappa a_i \overline q'^i\, .
\end{equation}

Solving the constraint equations $\Gamma_{i}(\sigma)=0$ and
$\Gamma(\sigma)=0$, we can express the closed string variables in terms of
the open string ones
\begin{equation}\label{eq:solution1}
\pi_{i}=p_{i} \,  , \quad x^{i}(\sigma) =q^i(\sigma)-2\int^{\sigma}_0 d \sigma_{1} \Big[
\Theta^{ij} p_j(\sigma_1)+\Theta^i p(\sigma_1) \Big] \,  ,
\end{equation}
\begin{equation}\label{eq:solution2}
\pi=p \, , \quad F(\sigma)=f(\sigma)+2\Theta^i\int^{\sigma}_0 d\sigma_1 p_i(\sigma_1)\, ,
\end{equation}
where the parameters are defined as
\begin{equation}
\Theta^{i j}=-\frac{1}{\kappa}(\tilde P_T B P_T^0)^{ij}=-\frac{1}{\k}(G_{eff}^{-1}\Pi_T^0 B G^{-1}\Pi_T^0)^{i j} \,  , \qquad
\Theta^{i}=\frac{(aB\tilde G^{-1})^i}{2\k a^2}=\frac{(\tilde a
BG^{-1})^i}{2\k \tilde a^2} \,  ,
\end{equation}
with 
\begin{equation}\label{eq:tildep}
(\tilde P_T)^{ij}=(G_{eff}^{-1})^{ik}\left[ \delta_k{}^j-(P_1)_k{}^j-(\Pi_0)_k{}^j\right]\, .
\end{equation}
The projectors $(\Pi_T^0)_i{}^j$ and $(P_1)_i{}^j$ are introduced in (\ref{eq:pitenula}) and (\ref{eq:pet1}).

The noncommutativity tensor $\Theta^{ij}$ has a clear geometrical interpretation. The induced close string metric $g_{ij}$ on the $p$ dimensional submanifold orthogonal to vector $a_i$ is $(P_T^0G)_{ij}\,$, while the corresponding open string one $\tilde g_{ij}$ is $(\tilde P_T)_{ij}$. Consequently, in
terms of the induced metrics on $D_{p-1}$-brane, the
noncommutativity parameter for $x^i$ variables has the same form as in the
dilaton free case
\begin{equation}\label{eq:tetka1}
\Theta^{ij}=-\frac{1}{\kappa}(\tilde g^{-1}B g^{-1})^{ij}\, .
\end{equation}

We are going now to find the effective theory in terms of the open string
variables. In analogy with the closed string currents
(\ref{eq:Jstruja})-(\ref{eq:jstruja}), we introduce open string currents
denoted by $\tilde i^\Phi_{\pm}$, $\tilde j_{\pm i}$, $\tilde j$ and $\tilde
J^i_\pm$ substituting the closed string variables and metric tensor with open
ones and omitting the antisymmetric field. Correlating the closed string
currents with the open string currents, we obtain
\begin{equation}
T_{\pm}= \tilde T_{\pm} \, , \quad \mathcal{H}_{c}=\tilde
\mathcal{H}_{c}  \,  ,
\end{equation}
where we introduced effective energy momentum tensor and Hamiltonian
\begin{equation}
\tilde T_{\pm}=\mp\frac{1}{4\kappa}\left [ (\tilde G^{-1})^{ij}\tilde J_{\pm
i}\tilde J_{\pm j}+\frac{\tilde j}{\tilde a^2}\tilde i^\Phi_\pm\right
]+\frac{1}{2}\left ( \tilde i'^\Phi_\pm-f'\tilde i^\Phi_\pm \right ) \, ,
\quad \tilde \mathcal{H}_{c}=\tilde T_- - \tilde T_+  \,  .
\end{equation}

Therefore, the effective theory has the same form as the original one, but in
terms of the symmetric variables $q^i$, $f$ and corresponding momenta $p_i$,
$p$ in new background $G_{ij}\to \tilde {G}_{ij}$, $B_{ij}\to 0$ and
$\Phi\to \Phi_0+a_i q^i$.

Using Poisson brackets of the closed string variables, we can calculate
the algebra of the open string ones
\begin{equation}\label{eq:pzagrada}
\{q^{i}(\tau,\sigma),p_{j}(\tau,\overline{\sigma})\}=
{\delta^{i}}_{j}\delta_{s}(\sigma,\overline{\sigma}) \,  ,\quad
\{f(\tau,\sigma),p(\tau,\overline{\sigma})\}=\delta_{s}(\sigma,\overline{\sigma})\,
,
\end{equation}
where the symmetric delta function is
\begin{equation}\label{eq:simdel}
\delta_{s}(\sigma,\overline\sigma)=\frac{1}{2}[\delta(\sigma-\overline\sigma)+
\delta(\sigma+\overline\sigma)].
\end{equation}
Now, it is easy to calculate Poisson brackets
\begin{equation}
\{ x^i (\sigma) , x^j ({\bar \sigma}) \} = 2 \Theta^{i j}
\theta(\sigma + {\bar \sigma})  \,  ,
\end{equation}
\begin{equation}
\{ x^i (\sigma) , F ({\bar \sigma}) \} = 2 \Theta^i \theta(\sigma
+ {\bar \sigma})  \,  , \qquad  \{ F(\sigma) , F ({\bar \sigma}) \} = 0  \,  ,
\end{equation}
where the function $\theta(x)$ is defined as
\begin{equation}\label{eq:fdelt}
\theta(x)=\left\{\begin{array}{ll}
0 & \textrm{if $x=0$}\\
1/2 & \textrm{if $0<x<2\pi$}\, .\\
1 & \textrm{if $x=2\pi$} \end{array}\right .
\end{equation}

It is useful to separate the center of mass closed string coordinates
\begin{equation}\label{eq:cenmassX}
x^i_{cm}=\frac{1}{\pi}\int_0^\pi d\sigma x^i(\sigma)\, ,\quad
x^i(\sigma)=x^i_{cm}+X^i(\sigma)\, ,
\end{equation}
\begin{equation}\label{eq:cenmassF}
F_{cm}=\frac{1}{\pi}\int_0^\pi d\sigma F(\sigma)\, ,\quad F(\sigma)=
F_{cm}+\mathcal F(\sigma)\, ,
\end{equation}
so that we have
\begin{equation}
\{ X^i(\sigma), X^j(\overline \sigma) \}=\Theta^{ij}\Delta(\sigma+\overline
\sigma)\, ,
\end{equation}
\begin{equation}
\{ X^i(\sigma), \mathcal F(\overline \sigma)
\}=\Theta^{i}\Delta(\sigma+\overline \sigma)\, ,\quad \{ \mathcal F(\sigma),
\mathcal F(\overline \sigma) \}=0\, ,
\end{equation}
where
\begin{equation}\label{eq:DELTA}
\Delta(x)=2\theta(x)-1=\left \{
\begin{array}{ll}
-1 & \textrm{if $x=0$}\\
0 & \textrm{if $0<x<2\pi$}\, .\\
1 & \textrm{if $x=2\pi$}
\end{array}\right.
\end{equation}
Therefore, the variables $X^i$ and $\mathcal F$ are commutative in the interior of string, while on the string
endpoints they are noncommutative.

As a consequence of the relations $a_i \Theta^{i j} = 0$ and $a_i \Theta^i =
0$, coordinate $x_0$ [see eq.(\ref{eq:notacija})] becomes a commutative one. On the other hand, $F$ is
the new noncommutative variable and the number of the noncommutative variables
is the same as in the dilaton free case.

\section{Noncommutativity in the case $a^2=0$ and $\tilde a^2\neq0$}

\setcounter{equation}{0}

In simple dynamical theories velocities (time derivatives of the coordinates)
can be expressed in terms of the canonical momenta. However, when the coefficient in
front of velocity is equal to zero, there appears a constraint in the theory (see for instance ref.\cite{BN}).

Here we are going to investigate a particular case when the gradient of dilaton
field $a_i$ is light-like vector with respect to the closed string metric
$G_{ij}$, $a^2=0$. This condition causes the existence of a first class
constraint.

\subsection{Canonical analysis}

The canonical momenta conjugated to $x^i$ and $F$
\begin{equation}
\pi_i=\kappa (G_{ij}\dot x^j-2 B_{ij}x'^j+2a_i \dot F)\, ,\quad \pi=2\kappa a_i \dot x^i\, ,
\end{equation}
can be combined as
\begin{equation}
j \equiv a^i \pi_i-\frac{1}{2}\pi +2\kappa a^i B_{i j} {x^j}' = 2\kappa a^2 \dot F\, ,
\end{equation}
which means that, for $a^2=0$, $j$ is a constraint of the theory.

The canonical Hamiltonian obtained by standard definition
$\mathcal{H}_c=\pi_i \dot x^i+\pi\dot F-\mathcal{L}$
\begin{equation}\label{eq:hami}
\mathcal{H}_c=\frac{1}{2\k} \pi_i G^{ij} \pi_j + \frac{\k}{2} {x^i}' G^{eff}_{ij} {x^j}' +
2 \pi_i (G^{-1}B)^i_j {x^j}'+ 2\k  a_i{x^i}'F'  \,  ,
\end{equation}
can be expressed in terms of currents
\begin{equation}\label{eq:zvezdaj}
j^*_{\pm i}=j_{\pm i}\pm 2\kappa a_i F'\, ,
\end{equation}
in the form
\begin{equation}\label{eq:hamilton}
\mathcal{H}_c=T_- - T_+ =\frac{1}{4\k} G^{ij} (j^*_{- i}j^*_{- j}+ j^*_{+ i}j^*_{+ j})
 \,  .
\end{equation}
The constraint $j$ has the same form in terms of the Dp-brane currents (\ref{eq:jistruja}) as in terms of the currents $j_{\pm i}^*$
\begin{equation}\label{eq:veza}
j=a^i j_{\pm i}-\frac{1}{2}  i^\Phi_{\pm} =a^i j_{\pm i}^*-\frac{1}{2}i^\Phi_{\pm}   \, .
\end{equation}

To investigate the theory with constraints, we introduce total
Hamiltonian
\begin{equation}\label{eq:totalni}
H_T=\int d\sigma \mathcal{H}_T\, ,\quad \mathcal{H}_T=\mathcal{H}_c+\lambda j  \,  ,
\end{equation}
where  $\lambda$ is a Lagrange multiplier.

Using the Poisson brackets of the basic canonical variables, we obtain
\begin{equation}
\{j^*_{\pm i},j^*_{\pm j}\}=\pm2\k G_{ij}\delta'  \,  , \qquad
\{j^*_{\pm i},i^\Phi_{\pm}\}=\pm 4\k a_i \delta'   \,  ,
\end{equation}
while all opposite chirality currents commute. It is easy to show that
\begin{equation}
\{j^*_{\pm i},j\}=0  \,  ,  \qquad  \{ H_T,j\}=0   \,  ,
\end{equation}
and consequently $j$ is a first class constraint. So, the theory is invariant
under some local symmetry.

The gauge transformation of any variable $X$ is defined in terms of symmetry
generator $G$ in the form
\begin{equation}
\delta_{\eta} X= \{X,G \}\, ,  \qquad   G \equiv \int d \sigma \eta
(\sigma) j (\sigma)  \,  ,
\end{equation}
which produces
\begin{equation}\label{eq:prom1}
\delta_{\eta}x^i=a^i \eta \,   ,  \quad
\delta_{\eta}F=-\frac{1}{2}\eta  \,  ,\qquad \delta_{\eta}\pi_i=2\kappa a^j B_{j i} \eta' \,  ,  \quad
\delta_{\eta}\pi=0  \,  .
\end{equation}
The transformation of the intrinsic metric tensor has form
\begin{equation}
\delta_{\eta} g_{\alpha\beta}=-\eta g_{\alpha\beta}  \,  ,
\end{equation}
and we recognize the two dimensional Weyl symmetry as a part of the above
gauge symmetry.

\subsection{Gauge fixing and solution of constraints}

The gauge freedom allows us to fix one degree of freedom. We choose $F=0$ and after that we can treat $j$ and $F$ as second class constraints. On the constraints we have
\begin{equation}
J_{\pm i}\to j_{\pm i}\, ,\quad i^F_\pm\to 0\, ,\quad i^\Phi_\pm\to 2 a^i j_{\pm i}\, ,
\end{equation}
so that the boundary conditions (\ref{eq:gruslov}) take the form
\begin{equation}\label{eq:novigruslov}
\gamma_i^{(0)}=\Pi_{+\,i}{}^j j_{-\,j}+\Pi_{-\,i}{}^j j_{+\,j}\, ,\quad \gamma^{(0)}=a^i(j_{-\,i}-j_{+\,i})\, .
\end{equation}
They are considered as canonical constraints. Examinig the consistency of the constraints, we obtain
\begin{equation}
\Gamma_i(\sigma)=\Pi_{+ i}{}^j j_{- j}(\sigma)+\Pi_{- i}{}^j j_{+ j}(-\sigma)\, ,\quad \Gamma(\sigma)=a^i\left[ j_{- i}(\sigma)-j_{+ i}(-\sigma)\right]\, ,
\end{equation}
which satisfy algebra
\begin{equation}
\left\lbrace \Gamma_i(\sigma), \Gamma_j(\overline\sigma)\right\rbrace=-\kappa G^{eff}_{ij}\delta'\, ,\quad \left\lbrace  \Gamma_i(\sigma), \Gamma(\overline\sigma)\right\rbrace=-2\kappa a_i \delta'\, ,\quad \left\lbrace  \Gamma(\sigma), \Gamma(\overline\sigma)\right\rbrace=0\, .
\end{equation}
The Poisson brackets of the complete set of the constraints
$\chi_A=(\Gamma_i,\Gamma)$ can be written in a matrix form
\begin{equation}\label{eq:IIklasa}
\left\lbrace \chi_A(\sigma), \chi_B(\overline \sigma) \right\rbrace=-\kappa M_{AB}\delta'\, ,
\end{equation}
where
\begin{equation}\label{eq:matM2}
M_{AB}=\left (
\begin{array}{rr}
G^{eff}_{ij} & 2a_i\\
2a_j & 0
\end{array}\right )\, .
\end{equation}
We assume that $\det G^{eff}_{ij}\neq0$ and with the help of the relation
\begin{equation}
\det M_{AB}=-4\tilde a^2 \det G^{eff}_{ij}\, ,
\end{equation}
we conclude that for $\tilde a^2\neq0$ all constraints originated
from boundary conditions are of the second class.

In terms of the open string variables introduced in
(\ref{eq:mena1})-(\ref{eq:mena2}), the constraint equations, $\Gamma_i(\sigma)=0$ and $\Gamma(\sigma)=0$, produce
\begin{equation}\label{eq:prvagrupa}
\overline p_i=0\, ,\quad \overline q'^i=\frac{2}{\kappa}(G_{eff}^{-1}BG^{-1})^{ij}p_j\, ,
\end{equation}
\begin{equation}\label{eq:drugagrupa}
a_i \overline q'^{i}=0\, ,\quad a^i \overline p_i+2\kappa (aB)_i q'^i=0\, .
\end{equation}
Treating gauge fixing and the first class
constraints as second class constraints, from $j=0$ and $F=0$ we
obtain additional equations
\begin{equation}\label{eq:fccs}
a^i p_i-\frac{1}{2}p+2\kappa (aB)_i \overline q'^i=0\, ,\quad a^i \overline p_i-\frac{1}{2}\overline p+2\kappa (aB)_i q'^i=0\, ,\quad f=0\, ,\quad \overline f=0\, .
\end{equation}
From the first equation in (\ref{eq:prvagrupa}) and second in (\ref{eq:drugagrupa}) follows $(aB)_i q'^i=0$, which in notation of (\ref{eq:notacija}) can be rewritten as $q_1'=0$. The second equation in (\ref{eq:prvagrupa}) and the first equation in (\ref{eq:drugagrupa}) give $(\tilde aB)^i p_i=0$ i.e. $p_1=0$. Consequently, the string dynamics is desribed by the following coordinates and momenta
\begin{equation}
(q_T^1)^i=(P_T^1)^i{}_j q^j\equiv Q^i\,\quad (\pi_T^1)_i=(P_T^1)_i{}^j p_j\equiv P_i\, ,
\end{equation}
where the projector $P_T^1$ is  defined in (\ref{eq:pet1}).
Now, we can rewrite (\ref{eq:prvagrupa}) in the form $\overline q'^i=\frac{2}{\kappa}(G_{eff}^{-1}BG^{-1}P_T^1)^{ij}P_j$. It is useful to separate it into direction along $(aB)_i$ and the orthogonal one $\overline q'^i=\overline q_1'^i+(\overline q_T^1)'^i$
\begin{equation}
(\overline q_T^1)'^i=-2\Theta^{ij}P_j\, ,\quad \overline q_1'^i=\frac{2}{\kappa}(G_{eff}^{-1}P_1 BG^{-1})^{ij}P_j\, ,
\end{equation}
so that tensor 
\begin{equation}
\Theta^{ij}=-\frac{1}{\kappa}( G_{eff}^{-1}P_T^1 B G^{-1}P_T^1)^{ij}\, ,
\end{equation}
is antisymmetric. Substituting expression for $\overline q'^i$ in the first equation (\ref{eq:fccs}), we have $p=2\tilde a^i P_i$.

The final solution of the equations (\ref{eq:prvagrupa})-(\ref{eq:fccs}) is
\begin{equation}\label{eq:drugoresenje}
x^i_{D_p}(\sigma)=Q^i(\sigma)-2\Theta^{ij}\int^{\sigma}_0 d\sigma_1 P_j(\sigma_1)\, ,\quad  \pi_i^{D_p}=P_i\, ,
\end{equation}
\begin{equation}\label{eq:x1pi1}
x_1(\sigma)=\frac{2}{\kappa}(\tilde aB^2G^{-1})^i\int^\sigma_0 d\sigma_1 P_i(\sigma_1)\, , \quad \pi_1=0\, ,
\end{equation}
\begin{equation}
F=0\, ,\quad \pi=2 \tilde a^i P_i\, ,
\end{equation}
where we choose integration constant $\overline q^i(\sigma=0)=0$. From (\ref{eq:x1pi1}) and periodicity of $P_i$ follows that $x_1$ satisfies Dirichlet boundary conditions i.e. $x_1(\sigma=0)=0=x_1(\sigma=\pi)$.

\subsection{Effective theory and noncommutativity}

Let us introduce effective current
\begin{equation}
\tilde j_{\pm i}=P_i\pm \kappa (P_T^1 G_{eff})_{ij}Q'^j\, ,
\end{equation}
and correlate it with $j_{\pm i}^*$ defined in (\ref{eq:zvezdaj})
\begin{equation}
j_{\pm i}^*=\pm 2 (\Pi_{\pm}G_{eff}^{-1})_i{}^j \tilde j_{\pm j}\, .
\end{equation}
Substituting this relation in (\ref{eq:hamilton}), we obtain the effective Hamiltonian
\begin{equation}
\tilde \mathcal H_c=\tilde T_{-}-\tilde T_{+}\, ,\quad \tilde
T_{\pm}=\mp\frac{1}{4\kappa}(G_{eff}^{-1}P_T^1)^{ij} \tilde j_{\pm
i}\tilde j_{\pm j}
\end{equation}

From definitions of the current $\tilde j_{\pm i}$ and the
expression for the energy-momentum tensor $\tilde T_{\pm}$ we can
conclude that the effective metric tensors are
\begin{equation}
g^{eff}_{ij} = (P_T^1 G^{eff})_{ij} \,  ,  \qquad   g_{eff}^{ij} = (G_{eff}^{-1} P_T^1)^{ij} \, .
\end{equation}
In fact they are induced metrics on subspace defined by the projector $P_T^1$. This projector also plays the role of unity in this subspace $g^{eff}_{ik}
g_{eff}^{kj} = (P_T^1)_i{}^j$.

Effective theory is expressed in terms of the open string variables, $Q^i$ and $P_i$, which satisfy the algebra
\begin{equation}
\left\lbrace Q^i(\sigma),P_j(\overline\sigma)\right\rbrace=(P_T^1)^i{}_j \delta(\sigma-\overline\sigma)\, ,
\end{equation}
in the background
\begin{equation}
G_{ij}\to g^{eff}_{ij}\, ,\quad B_{ij}\to
0\, ,\quad \Phi\to 0\, .
\end{equation}

Using the solution (\ref{eq:drugoresenje}) we find the
noncommutativity relation
\begin{equation}
\left\lbrace x^i_{D_p}(\tau,\sigma),x^j_{D_p}(\tau,\overline
\sigma)\right\rbrace = 2\Theta^{i j}\theta(\sigma+\overline
\sigma)  \,  ,
\end{equation}
where the function $\theta(x)$
is defined in eq.(\ref{eq:fdelt}). If we separate the center of mass variables  in the way described in
(\ref{eq:cenmassX})-(\ref{eq:cenmassF}), $x^i_{D_p}(\sigma)=(x^i_{D_p})_{cm}+X_{D_p}^i(\sigma)$,
we obtain
\begin{equation}
\{ X_{D_p}^i(\sigma), X_{D_p}^j(\overline \sigma)\}=\Theta^{ij}\Delta(\sigma+\overline \sigma)\, ,
\end{equation}
where the function $\Delta(x)$ is defined in (\ref{eq:DELTA}).

The solution (\ref{eq:x1pi1}) for the closed string coordinate $x_1$ satisfies Dirichlet boundary conditions, while the conformal part of the intrinsic metric $F$ satisfies these boundary conditions automatically, because $F=0$.
So, the number of Dp-brane dimensions decreases from $p+2$ to $p$. On the other
hand, we have $a_i Q^i\neq0$ and $a_i \Theta^{ij}=0$, so that
$x_c=a_i x^i_{D_p}=a_i Q^i$ is commutative coordinate, because it
is momentum independent variable. There are one commutative
coordinate, in the direction of the dilaton gradient $a_i$, and
$p-1$ noncommutative ones.

\section{Noncommutativity in the case $\tilde a^2=0$ and $a^2\neq0$}
\setcounter{equation}{0}

As a consequence of (\ref{eq:dklasa}), the singularity of the matrix $M_{AB}$
turns some constraints into the first class. According to the (\ref{eq:detM})
the singularity corresponds to the case $\tilde a^2=0$. In this section we also suppose
$a^2\neq0$.

In the case we are going to deal with, not only  $M_{AB}$ but also $\tilde
G_{ij}=(\hat P_T^1 G_{eff})_{ij}$ is singular. For $\tilde a^2=0$, $\hat P_T^1$ is the projector on the subspace orthogonal to the vector $(\tilde aB)^i$ [see eq.(\ref{eq:kapapet1})], which can be expressed as $(\tilde aB G^{-1}\tilde G)_i=0$. So, $\det M_{AB}$, as a function of
$\tilde a^2$, has two zeros in $\tilde a^2=0$ and we can expect to have two
first class constraints.

The canonical analysis, boundary conditions and the related consistency procedure
are the same as in the case $a^2\neq0$ and $\tilde a^2\neq0$, of Sec.3. Let us
turn to the separation of the first and the second class constraints.

\subsection{Classification of the constraints}

According to the canonical procedure for the constrained systems, we introduce total Hamiltonian
\begin{eqnarray}
H_T=\int d\sigma \mathcal{H}_T \, ,\quad
\mathcal{H}_T=\mathcal{H}_c+\lambda^i(\sigma)\Gamma_i(\sigma)+\lambda(\sigma) \Gamma(\sigma)\, ,
\end{eqnarray}
where $\mathcal H_c$, $\Gamma_i$ and $\Gamma$ are the canonical Hamiltonian and constraints defined in (\ref{eq:hamiltonijan}), (\ref{eq:velikog}) and (\ref{eq:malog}), while
$\lambda^i$ and $\lambda$ are Lagrange multipliers. The consistency conditions
$\{H_T,\Gamma_i(\sigma)\}\approx0$ and $\{H_T,\Gamma(\sigma)\}\approx0$
produce
\begin{eqnarray}\label{eq:jedn3}
{\Gamma'}_i\approx -\k \tilde G_{ij}{\lambda'}^j- 2\k a_i
\lambda' \, ,\quad {\Gamma'}\approx-2\k a_i {\lambda'}^i\,.
\end{eqnarray}
Here "$\approx$" is a symbol for weak equality, which means that it is fulfilled on the constraints.

With the help of the projectors $(P_0)_i{}^j$, $(\hat P_1)_i{}^j$,  and $(\hat P_T)_i{}^j$, defined in (\ref{eq:ptpl}), (\ref{eq:kapapet1})  and (\ref{eq:pete1}) respectively, we decompose the Lagrange multipliers $\lambda ^i$
\begin{equation}\label{eq:razlaganje}
\lambda^i=(\lambda_T)^i+(\lambda_1)^i+(\lambda_0)^i=(\lambda_T)^i+2\Lambda_1(\tilde
aB)^i+\Lambda_2\tilde a^i\, ,
\end{equation}
where $\Lambda_1=-\frac{2}{a^2}(\lambda Ba)$ and
$\Lambda_2=\frac{(aG_{eff}\lambda)}{a^2}$. Now, the consistency conditions (\ref{eq:jedn3}) turn to the equations
\begin{equation}\label{eq:okof}
(\lambda+\frac{1}{2}\Lambda_2)'=-\frac{a^i\Gamma_i'}{2\kappa a^2}\, ,\quad
(\lambda_T')^i=-\frac{1}{\kappa}(G^{-1}_{eff})^{ij}(P_T^0)
_j{}^k\Gamma'_k\, ,
\end{equation}
where $P_T^0$ is the projector defined in (\ref{eq:ptpl}). So, the coefficients
$\Lambda_1$ and $\Lambda_2$ are not determined.

If we rewrite the term $\lambda^i \Gamma_i$ using the equation
(\ref{eq:razlaganje}), we obtain
\begin{equation}\label{eq:popham}
H_T=H_c+\int_0^\pi d\sigma \left[ (\lambda_T)^i( \Gamma_T)_i+(\lambda+\frac{1}{2}\Lambda_2) \Gamma+\Lambda_1 \Gamma_1+\Lambda_2 \Gamma_2\right] =H'+\int_0^\pi d\sigma (\Lambda_1 \Gamma_1+\Lambda_2 \Gamma_2)\, ,
\end{equation}
where
\begin{equation}\label{eq:prvi}
\Gamma_1=2(\tilde a BG^{-1})^i \Gamma_i\, ,\quad \Gamma_2=\tilde
a^i \Gamma_i-\frac{1}{2}\Gamma\, .
\end{equation}
The constraints, $\Gamma_1$ and $\Gamma_2$, multiplied by the arbitrary
coefficients $\Lambda_1$ and $\Lambda_2$, are of the first class, while $(\Gamma_T)_i$ and $\Gamma$, multiplied by the determined multipliers, are of the second class. 

Using the algebra of the constraints (\ref{eq:dklasa}), we can easily calculate
\begin{equation}\label{eq:nule1}
\left\lbrace \Gamma_1,\Gamma_i\right\rbrace =0\, ,\quad \left\lbrace
\Gamma_1,\Gamma\right\rbrace =0\, , \qquad \left\lbrace
\Gamma_2,\Gamma_i\right\rbrace =0\, ,\quad \left\lbrace
\Gamma_2,\Gamma\right\rbrace =0\, ,
\end{equation}
and confirm the result that constraints $\Gamma_1$ and $\Gamma_2$
are of the first class.

Let us mark the set of second class constraints with $\chi_A=\left\lbrace (\Gamma_T)_i, \Gamma \right\rbrace $ and define the matrix $M_{AB}$ with
the relation $\left\lbrace \chi_A, \chi_B \right\rbrace=-\kappa
M_{AB}\delta' $.  We easily obtain
\begin{equation}
M_{AB}=\left(
\begin{array}{rr}
M_{ij} & 2 a_i \\ 2 a_j & 0
\end{array}  \right)\, ,
\end{equation}
where $M_{ij}= (\hat P_T G_{eff})_{ij}-(P_0 G_{eff})_{ji}$. The
matrix $M_{AB}$ maps two vectors to zero
\begin{equation}\label{eq:singsing}
M\left(
\begin{array}{c}
\tilde a \\ 0
\end{array}\right)=0\, ,\quad M\left(
\begin{array}{c}
(\tilde a B)\\0
\end{array}\right)=0\, .
\end{equation}
So, its rank is not greater then $p$. We suppose that the rest
of the matrix $M_{AB}$ is regular.

\subsection{Gauge fixing and solution of the constraints}

The first class constraints generate a local symmetry in the form
\begin{equation}\label{eq:localsym}
\delta X=\left\lbrace X,G\right\rbrace \, ,\quad G=\int_0^\pi d\sigma (\eta_1\Gamma_1+\eta_2\Gamma_2)\, ,
\end{equation}
where $\eta_1$ and $\eta_2$ are the parameters of the transformations. The expressions (\ref{eq:prvi}) in
terms of the open string variables have form
\begin{equation}\label{eq:fc}
\Gamma_1=\tilde a^i p_i-\frac{1}{2}p+2(\tilde aBG^{-1})^i\overline p_i\,
,\quad \Gamma_2=\tilde a^i \overline p_i-\frac{1}{2}\overline p+2(\tilde
aBG^{-1})^i p_i\, .
\end{equation}

They generate gauge transformations
\begin{equation}\label{eq:g1}
\delta q^i=\tilde a^i(\eta_1)_s+2(\tilde aBG^{-1})^i(\eta_2)_s\, ,\quad \delta
f=-\frac{1}{2}(\eta_1)_s\, ,
\end{equation}
\begin{equation}\label{eq:g2}
\delta \overline q^i=\tilde a^i(\eta_2)_a+2(\tilde aBG^{-1})^i(\eta_1)_a\,
,\quad \delta \overline f=-\frac{1}{2}(\eta_2)_a\, ,
\end{equation}
where the indices $"s"$ and $"a"$ denote $\sigma$ symmetric and antisymmetric
parts of the parameters $\eta_1$ and $\eta_2$. The gauge transformations of the coordinate components are of the forms
\begin{equation}
\delta (q_1)^i=2(\tilde a BG^{-1})^i\eta_{2s}\, ,\quad \delta
(q_0)^i=\tilde a^i\eta_{1s}\, ,\quad \delta (q_T)^i=0\,
,
\end{equation}
\begin{equation}
\delta (\overline q_1)^i=2(\tilde aBG^{-1})^i\eta_{1a}\,
,\quad \delta (\overline q_0)^i=\tilde a^i\eta_{2a}\, ,\quad
\delta (\overline q_T)^i=0\, .
\end{equation}

To fix the gauge corresponding to the parameters $\eta_{2s}$,
$\eta_{1a}$, $\eta_{1s}$ and $\eta_{2a}$, we choose
\begin{equation}\label{eq:gejdz}
(q_1)^i=0\, ,\quad (\overline q_1)^i=0\, ,
\quad f=0\, ,\quad \overline f=0\, ,
\end{equation}
which are equivalent to
\begin{equation}\label{eq:gauge}
(aB)_i q^i=0\, ,\quad (aB)_i \overline q^i=0\, ,\quad f=0\, ,\quad
\overline f=0\, .
\end{equation}

The set of the first class constraints and gauge conditions behave like a
set of the second class constraints. The original second class constraints together with first class constraints (\ref{eq:fc}) cause that full expressions $\Gamma_i$ and $\Gamma$ vanish
\begin{equation}\label{eq:2klasa}
2{(BP_T^0)_i}^j p_j+\frac{1}{a^2}(Ba)_i p-\k
\tilde G_{ij}{\overline q'}^j-2\k a_i {\overline f}'=0\, ,\quad \overline p_i=0\, ,\quad \overline p=0\, ,\quad a_i \overline q'^i=0\, .
\end{equation}

The complete set of the equations consists of (\ref{eq:gejdz}), (\ref{eq:2klasa}) and remained consequences of the first class constraints
\begin{equation}
(\tilde aB)^i p_i=0\, ,\quad p=2\tilde a^i p_i\, .  
\end{equation}
Choosing integration constant $\overline q^i(\sigma=0)=0$ we obtain solution 
\begin{equation}\label{eq:resenje12}
x^i_{D_p}(\sigma)=\hat Q^i(\sigma)-2\Theta^{ij}\int^\sigma_0 d\sigma_1 \hat P_j(\sigma_1)\,
,\quad \pi_i^{D_p}=\hat P_i \,  ,\qquad x_1=0\, ,\quad \pi_1=0\, , 
\end{equation}
\begin{equation}\label{eq:Fresenje}
F=0\, ,\quad
\pi=2\tilde a^i \hat P_i\, ,
\end{equation}
expressed in terms of the open string variables 
\begin{equation}
(q_T^1)^i=(\hat P_T^1)^i{}_j q^j\equiv \hat Q^i\, ,\quad (p_T^1)_i=(\hat P_T^1)_i{}^j p_j\equiv \hat P_i\, .
\end{equation}

The tensor $\Theta^{ij}$
\begin{equation}\label{eq:teta5}
\Theta^{ij}=-\frac{1}{\kappa}(G_{eff}^{-1}\hat P_T^1 B G^{-1}\hat P_T^1)^{ij}\, ,
\end{equation}
is manifestly antisymmetric.

\subsection{Effective theory}

Similarly as in subsection 4.3, we define the effective current
\begin{equation}
\tilde j_{\pm i}=\hat P_i \pm \kappa (\hat P_T^1 G_{eff})_{ij}\hat
Q'^j\, ,
\end{equation}
and correlate it with the currents $j^*_{\pm i}$ (\ref{eq:zvezdaj}) and
$i^\Phi_{\pm}$ (\ref{eq:jstruja})
\begin{equation}\label{eq:efekat2}
j_{\pm i}^*=\pm 2(\Pi_{\pm} G_{eff}^{-1})_i{}^j \tilde j_{\pm
j}-4(G_{eff}^{-1}\Pi_{\pm}\hat P_1 B)^{ij}p_j\, ,\quad
i^\Phi_{\pm}=2\tilde a^i\tilde j_{\pm i}\, .
\end{equation}

We are  going to express the Hamiltonian $H'$, defined in (\ref{eq:popham}), in terms of the open string variables. It is enough to do this for
canonical Hamiltonian because we already solved the second class constraints.

The energy-momentum components (\ref{eq:enmomten}) can be rewritten in terms of $j^*_{\pm i}$ and $i^\Phi_\pm$ as
\begin{equation}
T_{\pm}=\mp \frac{1}{4\kappa}\left[ G^{ij}j^*_{\pm i}j^*_{\pm j}- \frac{(a^i
j^*_{\pm i})^2}{a^2}+\frac{a^ij^*_{\pm
i}}{a^2}i^\Phi_{\pm}-\frac{1}{4a^2}i^\Phi_{\pm}i^\Phi_{\pm}\right] \, .
\end{equation}

Substituting the relations (\ref{eq:efekat2}) in the expression for
energy-momentum tensor we obtain
\begin{equation}
T_{\pm}=\tilde T_{\pm}\, ,\quad \tilde \mathcal{H}_c=\tilde T_{-}-\tilde T_{+}\equiv \mathcal H_c\, ,
\end{equation}
where
\begin{equation}
\tilde T_{\pm}=\mp \frac{1}{4\kappa}(G_{eff}^{-1}\hat P_T^1)^{ij}\tilde j_{\pm
i}\tilde j_{\pm j}\, .
\end{equation}

In terms of effective variables $\hat Q^i$ and $\hat P_j$, the effective
theory lives in the background $G_{ij} \to g^{eff}_{ij}=(\hat P_T^1 G_{eff})_{ij}, \, B_{ij} \to 0 , \, \Phi
\to 0$, and does not explicitly depend on antisymmetric and dilaton fields.

\subsection{Noncommutativity}

Using the algebra of the variables $q^i$ and $p_j$ (\ref{eq:pzagrada}), we can
calculate the Poisson brackets of the open string variables $\hat Q^i$ and
$\hat P_j$
\begin{equation}
\left\lbrace \hat Q^i(\sigma), \hat P_j(\overline \sigma) \right\rbrace=
(\hat P_T^1)^i_{\;j} \delta_s(\sigma,\overline
\sigma)\, ,
\end{equation}
where $\delta_s(\sigma,\overline \sigma)$ is defined in equation
(\ref{eq:simdel}).

As in the two previous cases, with the help of the equations (\ref{eq:resenje12}), after separation of the center
of mass variables, we obtain
\begin{equation}
\{X^{i}_{D_p}(\tau,\sigma),X^{j}_{D_p}(\tau,\overline{\sigma})\}=
\Theta^{ij}\Delta(\sigma+\overline{\sigma})\, ,
\end{equation}
where antisymmetric tensor $\Theta^{ij}$ and the function $\Delta(x)$ are
defined in (\ref{eq:teta5}) and (\ref{eq:DELTA}), respectively. We can easily conclude that the
interior of the string is commutative.

Let us discuss more precisely the noncommutativity of the string
endpoints. From (\ref{eq:resenje12}) and (\ref{eq:Fresenje}) we conclude that $x_1$ and $F$ satisfy Dirichlet boundary conditions and decrease
the number of Dp-brane dimensions from $p+2$ to $p$. The
projection on the Dp-brane is realized by operator $\hat
P_T^1$, (\ref{eq:kapapet1}). Similarly as in the subsection 4.3, we have $a_i \hat Q^i\neq0$ but
$a_i \Theta^{ij}=0$ so that $x_c \equiv a_i x^i_{D_p}=a_i \hat
Q^i$ does not depend on the momenta and, consequently, it is
commutative Dp-brane coordinate. All other $p-1$ Dp-brane
coordinates are noncommutative.

\section{Concluding remarks}
\setcounter{equation}{0}

In this paper we have considered the contribution of the linear dilaton field to
noncommutativity in some specific cases when the dilaton gradient $a_i$ is a
light-like vector either with respect to the closed string metric, $a^2=0$, or to the
open string one, $\tilde a^2=0$. We analyzed three cases: 1. $a^2\neq0$ and $\tilde
a^2\neq0$, 2. $a^2=0$ and $\tilde a^2\neq0$ and 3. $\tilde a^2=0$ and $a^2\neq0$. In all cases we treat the conformal
part of the world sheet metric, $F$, as a dynamical variable.

The first case  has been considered in
ref.\cite{BS2}. The Dp-brane
coordinate, in the direction of the dilaton gradient $a_i$, is commutative and $F$
takes the role of a new noncommutative coordinate.
So, there is one commutative and $p+1$ noncommutative coordinates.

In the second case, the condition $a^2=0$ produces one first class constraint.
In the third case, as a consequence of the condition $\tilde a^2=0$, two second class constraints turn to the first ones. In both cases, the first class constraints generate local
symmetries and after gauge fixing, the gauge conditions and the first class constraints can be treated as second class constraints. Solving them together with the original second class constraints we obtain effective coordinates and momenta describing the string. In both cases, Dp-brane is p-dimensional with one commutative coordinate, in the direction of the dilaton gradient $a_i$, and $p-1$ noncommutative coordinates.

Let us discuss general features of the solutions (\ref{eq:solution1})-(\ref{eq:solution2}),
(\ref{eq:drugoresenje})  and (\ref{eq:resenje12})
\begin{equation}
x^i_{D_p}(\sigma)=Q^i(\sigma)-2\Theta^{ij}\int^\sigma_0 d\sigma_1
P_{j}(\sigma_1)\, ,
\end{equation}
expressing closed string variables
$x^i_{D_p}=(P_{D_p})^i{}_j x^j$ in terms of open string ones and corresponding momenta 
\begin{equation}
Q^i=(P_{D_p})^i{}_j q^j\, ,\quad P_{i}=(P_{D_p})_i{}^j p_j\, .
\end{equation}
The variables $Q^i$ and $P_i$
satisfy the algebra
\begin{equation}
\left\lbrace
Q^i(\sigma),P_j(\overline\sigma)\right\rbrace=(P_{D_p})^i{}_j\delta_s(\sigma,\overline\sigma)\,
,
\end{equation}
where $P_{D_p}$ is unity in the Dp-brane subspace (see Table \textbf{I}).

Some closed string components
vanish on the solution of constraints, because the projections of both
the open string coordinates  and noncommutativity parameter vanish. It means that
corresponding components are fixed and effectively satisfy Dirichlet boundary
conditions. The other components, as $x_1$ in the second case, are different from zero, but also satisfies Dirichlet boundary conditions. Consequently, in both cases the number of Dp-brane dimensions decreases.
In fact, first class constraints turn some Neuman to Dirichlet boundary conditions, so
that, after carefully carried calculations, the true dimension of the Dp-brane is $D_{Dp}=2p+4-2N_{FCC}-N_{SCC}$, where $N_{FCC}$ is the number of the first class constraints and $N_{SCC}$ the number of the second class ones.

If the closed string components contain only the open string
coordinates, while the momenta are absent, they are commutative
degrees of freedom. In all cases, in a direction defined by the
vector $a_i$, the noncommutativity parameter $\Theta^{ij}$ is
singular, $a_i \Theta^{ij}=0$, so the effective momentum
disappears from $a_i x^i_{D_p}$.  The corresponding effective
coordinate is nonzero, $a_i \, Q^i\neq0$, and consequently, one commutative coordinate $x_c=a_i x^i_{D_p}$ appears.

The closed string components $(x_{nc})^i$,
containing both the open string coordinates  and momenta, are the
noncommutative degrees of freedom. The noncommutativity relation has the same form in all three cases
\begin{equation}\label{eq:PB}
\{ X^i_{D_p}(\tau,\sigma),X^j_{D_p}(\tau,\overline
\sigma)\}=\Theta^{ij}\Delta(\sigma+\overline\sigma)\, , \quad
\Big[ X^i_{D_p}(\sigma)=x^i_{D_p}(\sigma)-(x^i_{D_p})_{cm}\Big]\, .
\end{equation}
When $a_i$ is not light-like vector the parameter $\Theta^{ij}$  is given in (\ref{eq:tetka1}), while in the other two cases, when $a_i$ is light-like vector, it can be expressed in terms of projectors $P_{D_p}$
\begin{equation}
\Theta^{ij}=-\frac{1}{\kappa}(G_{eff}^{-1} P_{D_p}
BG^{-1}P_{D_p})^{ij}\, .
\end{equation}
The number of the noncommutative coordinates, $N_{nc}$, is difference of Dp-brane dimension and
the number of the commutative coordinates, $N_{nc}=D_{Dp}-1$. All results are presented in the table {\bf{I}}
\vspace{0.5cm}

\begin{tabular}{|c|c|c|c|c|c|c|c|c|c|}\hline
\textbf{Case} & $N_{FCC}$ & $N_{SCC}$ & $D_{Dp}$ & $(P_{D_p})_i{}^j$ & VDbc & $(x_{nc})^i$ & $x_c$ & $g_{ij}^{eff}$\\ \hline\hline $\tilde a^2\neq0$, $a^2\neq0$ & 0 & p+2 & p+2 & $\delta_i{}^j$ & - & $(\Pi_T^0 x)^i\, ,F$ & $x_0$ & $\tilde G_{ij}$\\ \hline $a^2=0\, ,\tilde a^2\neq0$ & 1 & p+2 & p &
$(P_T^1)_i{}^j$ & $x_1\, ,F$ & $(P_T x)^i$ & $x_0$ & $(P_T^1 G_{eff})_{ij}$ \\ \hline $\tilde a^2=0\, ,a^2\neq0$ & 2
& p & p & $(\hat P_T^1)_i{}^j$ & $x_1\, ,F$ & $(\hat P_T x)^i$ & $x_0$ & $(\hat P_T^1 G_{eff})_{ij}$\\
\hline
\end{tabular}
\vspace{0.3cm}\\
where {\bf{VDbc}} means variables with Dirichlet boundary condition. The projectors $\Pi_T^0$,$P_T$, $\hat P_T$, $P_T^1$ and $\hat P_T^1$  as well as the coordinate projections ($x_0\, ,x_1$) are defined in Appendix.

In all cases the effective energy-momentum tensor $\tilde T_{\pm}$ satisfy Virasoro algebra. Consequently, the effective theory is a string theory propagating in the open string background $G_{ij}\to (g_{eff})_{ij}$, $B_{ij}\to 0$. The dilaton field survives only in the first case, again as a linear dilaton, $\Phi=\Phi_0+a_i q^i$.

The initial string, to which we refer as closed, is an oriented
string. An effective one, on the solution of boundary conditions,
is unoriented because it is symmetric under $\sigma \to -\sigma$.
It is well known, that an unoriented string does not contain
explicitly Kalb-Ramond field, because the term
$\varepsilon^{\alpha\beta}B_{\mu\nu}\partial_\alpha x^\mu
\partial_{\beta}x^\nu$ is not invariant under transformation
$\sigma \to -\sigma$. It explains the fact that in all cases the antisymmetric field
disappears from the effective background, $B_{ij}\to 0$. It can
survive only as bilinear combination, which indeed occurs as a
part of the effective metric.

At the end let us discuss the contribution of the Liouville term which offers a new viewpoint of the results of this paper. It is known from ref.\cite{FCB} that $\beta^G_{\mu \nu}=0 =
\beta^B_{\mu \nu}$ produce c-number value of the third beta-function playing the role of Schwinger term, $\beta^\Phi = c$. In that
case for $D= 26$  the central charge, $c= 4 a^2$, breaks conformal
invariance. In order to cancel the conformal anomaly the
corresponding Wess-Zumino term can be added to the action (\ref{eq:2deo}). In the particular
case this is just Liouville action
\begin{equation}\label{eq:kinclan}
S_{L}=  \frac{2 \kappa}{\alpha} \int_{\Sigma} d^2 \xi
\eta^{\alpha\beta}\partial_\alpha F \partial_\beta F\, ,  \qquad
\left( \frac{2 \kappa}{\alpha}=  \frac{\hbar (D-26)}{48 \pi}
\right)
\end{equation}
which is kinetic term for conformal part of the metric, $F$. After
change the variables, $F \to {}^\star F = F + {\alpha \over 2} a_i
x^i$, the term with linear dilaton field disappears but at the
same time the space-time metric obtains new term $G_{i j} \to
{}^\star G_{i j}  = G_{i j} - \alpha a_i a_j$. It seems that
linear dilaton field does not change anyhting, because the new conformal
factor ${}^\star F$  decouples and we obtain the
standard action without dilaton field. But, it is not complete story, because new
space-time metric ${}^\star G_{i j}$ becomes singular and produces
the gauge symmetry in the theory. It is easy to check that for
$a^2 = { 1 \over \alpha}$ the vector $a^i$ is singular vector of
the metric ${}^\star G_{i j}$.

We carefully investigate all three cases when the Liouville action
is included. We find that all important results, such as the form of the solution
\begin{equation}
{}^\star x^i_{D_p}(\sigma)={}^\star Q^i(\sigma)-2{}^\star \Theta^{ij}\int^\sigma_0 d\sigma_1
{}^\star P_{j}(\sigma_1)\, ,
\end{equation}
\begin{equation}
{}^\star Q^i=({}^\star P_{D_p})^i{}_j q^j\, ,\quad {}^\star P_{i}=({}^\star P_{D_p})_i{}^j p_j\, , \quad \left\lbrace
{}^\star Q^i(\sigma),{}^\star P_j(\overline\sigma)\right\rbrace=({}^\star P_{D_p})^i{}_j\delta_s(\sigma,\overline\sigma)\,
,
\end{equation}
the dimensions of Dp-brane, the number of commutative and noncommutative variables and the noncommutativity relation
\begin{equation}\label{eq:PBL}
\{ {}^\star X^i_{D_p}(\tau,\sigma),{}^\star X^j_{D_p}(\tau,\overline
\sigma)\}={}^\star \Theta^{ij}\Delta(\sigma+\overline\sigma)\, , \quad
\Big[ {}^\star X^i_{D_p}(\sigma)={}^\star x^i_{D_p}(\sigma)-({}^\star x^i_{D_p})_{cm}\Big]\, ,
\end{equation}
are the same as in the absence of Liouville action. In both cases where $a_i$ is light-like vector either with respect to ${}^\star G_{ij}$ or ${}^\star G_{ij}^{eff}$, the noncommutativity parameters can be expressed as
\begin{equation}
{}^\star \Theta^{ij}=-\frac{1}{\kappa}(G_{eff}^{-1}\,{}^\star P_{D_p}
BG^{-1}\,{}^\star P_{D_p})^{ij}\, ,
\end{equation}
where projectors ${}^\star P_{D_p}$ are defined in the table {\bf{II}} and in the equations (\ref{eq:starpt}) and (\ref{eq:starptkapa}). In the first case, where $a_i$ is not light-like vector, the noncomutavity parameter takes the form
\begin{equation}
{}^\star \Theta^{ij}=-\frac{1}{\kappa}({}^\star G_{eff}^{-1}B\;{}^\star G^{-1})^{ij}=-\frac{1}{\kappa}(G_{eff}^{-1}\check \Pi_T^0 BG^{-1}\check \Pi_T^0)^{ij}\, ,
\end{equation}
where 
\begin{equation}
(\check \Pi_T^0)_i{}^j=(\Pi_T^0)_i{}^j+\frac{1}{1-\alpha \tilde a^2}(\Pi_0)_i{}^j\, ,
\end{equation}
and $\Pi_T^0$ and $\Pi_0$ are defined in (\ref{eq:pitenula}).

The only differences produced by Lioville term are that
local gauge symmetries appear for $a^2 = { 1 \over \alpha}$ and
$\tilde a^2 = { 1 \over \alpha}$ instead for $a^2=0$ and $\tilde
a^2 =0$ and that some variables change the roles , $ x_0 \to {}^\star F$ and  $F \to x_0$. So, instead of $x_0$, the variable ${}^\star F$ is commutative. In the first case $x_0$ is noncommutative instead of $F$ and in the second and third case $x_0$ satisfies the Dirichlet boundary conditions instead of $F$. Results of these investigations are presented in the table \textbf{II} and will be published separately. The projectors $({}^\star P_T)_i{}^j$ and $({}^\star \hat P_T)_i{}^j$ are defined in (\ref{eq:starpt}) and (\ref{eq:starptkapa}).\\
\vspace{0.5cm}

\begin{tabular}{|c|c|c|c|c|c|c|c|c|}\hline
\textbf{Case} & $N_{FCC}$ & $N_{SCC}$ & $D_{Dp}$ & $({}^\star P_{D_p})_i{}^j$ & VDbc & $({}^\star x_{nc})^i$ & ${}^\star x_c$\\ \hline\hline $\tilde a^2\neq\frac{1}{\alpha}$, $a^2\neq\frac{1}{\alpha}$ & 0 & p+2 & p+2 & $\delta_i{}^j$ & - & $x^i$ & ${}^\star F$  \\ \hline $a^2=\frac{1}{\alpha}\, ,\tilde a^2\neq\frac{1}{\alpha}$ & 1 & p+2 & p &
$({}^\star P_T)_i{}^j$ & $x_0\, ,x_1$ & $({}^\star P_T x)^i$ & ${}^\star F$ \\ \hline $\tilde a^2=\frac{1}{\alpha}\, ,a^2\neq\frac{1}{\alpha}$ & 2
& p & p & $({}^\star \hat P_T)_i{}^j$ & $x_0\, ,x_1$ & $({}^\star \hat P_T x)^i$ & ${}^\star F$\\
\hline
\end{tabular}
\vspace{0.3cm}\\

After quantization, Poisson brackets (\ref{eq:PB}) and (\ref{eq:PBL}) turn to the commutation relations. Now, we are ready to comment whether the conformal symmetry survives on the quantum level at the endpoints of the open string. The contribution of the field $F$ to the open string theory is nontrivial only in the first case, where it becomes a noncommutative variable instead of $x_0$. In other two cases it disappears (see Table {\bf{I}}). So, if in addition to space-time field equations (\ref{eq:betaG})-(\ref{eq:betaFi}), we require either $a^2=0$ or $\tilde a^2=0$, the open string theory will be conformally invariant. In the presence of Liouville term the corresponding variable ${}^\star F$ decouples from the theory and has the role of commutative coordinate (see Table {\bf{II}}). In that case the equations (\ref{eq:betaG})-(\ref{eq:betaB}) are enough for conformal invariance of the open string theory.

\appendix

\section{Projectors}
\setcounter{equation}{0}

In this appendix we introduce projector operators in order to separate noncommutative and nonphysical variables on Dp-brane as well as to express the noncommutativity parameter.

The projectors on the direction $n_i$ and on the subspace orthogonal to vector $n_i$ are 
\begin{equation}
(\Pi)_i{}^j=\frac{n_i n^j}{n^2}\, ,\quad (\Pi_T)_i{}^j=\delta_i{}^j-(\Pi)_i{}^j\, ,
\end{equation}
where $n^i=g^{ij}n_j$ and $n^2=n^i n_i$ . The transposed operator is 
\begin{equation}\label{eq:trans}
\Pi^i{}_j=g^{ik}\Pi_k{}^l g_{lj}\, ,
\end{equation}
and similarly for $\Pi_T$. Using these projectors we can decompose arbitrary covariant vectors  
\begin{equation}
\pi_i=(\pi_L)_i+(\pi_T)_i\, ,\quad (\pi_L)_i=(\Pi)_i{}^j \pi_j\, ,\quad (\pi_T)_i=(\Pi_T)_i{}^j \pi_j\, ,
\end{equation}
and contravariant ones
\begin{equation}
x^i=(x_L)^i+(x_T)^i\, ,\quad (x_L)^i=(\Pi)^i{}_j x^j\, ,\quad (x_T)^i=(\Pi_T)^i{}_j x^j\, .
\end{equation}

We are going to apply this procedure to few cases choosing particular vectors $n_i$ and metrics $g_{ij}$.

\subsection{Case $n_i=a_i$ and $g_{ij}=G_{ij}$}

For $n_i\to (n_0)_i=a_i$ and $g_{ij}\to G_{ij}$ we obtain 
\begin{equation}\label{eq:ptpl}
(\Pi)_i{}^j\to(P_0)_i{}^j=\frac{a_i a^j}{a^2}\, ,\quad (\Pi_T)_i{}^j\to(P_T^0)_i{}^j=\delta_i{}^j-(P_0)_i{}^j\, . 
\end{equation}

\subsection{Case $n_i=a_i$ and $g=G_{ij}^{eff}$}

For the same vector $n_i\to (n_0)_i=a_i$, but using the effective metric $g_{ij}\to G_{ij}^{eff}$, we have
\begin{equation}\label{eq:pitenula}
(\Pi)_i{}^j\to (\Pi_0)_i{}^j=\frac{a_i \tilde a^j}{\tilde a^2}\, ,\quad (\Pi_T)_i{}^j\to (\Pi_T^0)_i{}^j=\delta_i{}^j-(\Pi_0)_i{}^j\, .
\end{equation}

\subsection{Case $n_i=(aB)_i$ and $g_{ij}=G^{eff}_{ij}$}

Applying the same procedure, for vector $n_i\to (n_1)_i=(aB)_i$ and $g_{ij}\to G^{eff}_{ij}$, we have
\begin{equation}\label{eq:pipit1}
(\Pi)_i{}^j\to (\Pi_1)_i{}^j=\frac{4}{\tilde a^2-a^2}(Ba)_i(\tilde aB)^j\, ,\quad (\Pi_T)_i{}^j\to (\Pi_T^1)_i{}^j=\delta_i{}^j-(\Pi_1)_i{}^j\, .
\end{equation}

We introduce new notation for the cases where $a_i$ is light-like vector with respect to $G_{ij}$ ($a^2=0$)
\begin{equation}\label{eq:pet1}
(P_1)_i{}^j=(\Pi_1)_i{}^j\Big|_{a^2=0}=\frac{4}{\tilde a^2}(Ba)_i(\tilde aB)^j\, ,\quad (P_T^1)_i{}^j=(\Pi_T^1)_i{}^j\Big|_{a^2=0}=\delta_i{}^j-(P_1)_i{}^j\, ,
\end{equation}
and with respect to the effective metric $G_{ij}^{eff}$ ($\tilde a^2=0$)
\begin{equation}\label{eq:kapapet1}
(\hat P_1)_i{}^j=(\Pi_1)_i{}^j\Big|_{\tilde a^2=0}=-\frac{4}{a^2}(Ba)_i(\tilde aB)^j\, ,\quad (\hat P_T^1)_i{}^j=(\Pi_T^1)_i{}^j\Big|_{\tilde a^2=0}=\delta_i{}^j-(\hat P_1)_i{}^j\, .
\end{equation}

\subsection{Case $(n_0)_i=a_i$ and $(n_1)_i=(aB)_i$ and $g_{ij}=G^{eff}_{ij}$}

Let us construct the projector orthogonal to the vectors $(n_0)_i=a_i$ and $(n_1)_i=(aB)_i$ with respect to the effective metric $G_{ij}^{eff}$. These two vectors are mutually orthogonal and it is enough to use the constructed projectors on the direction $a_i$, (\ref{eq:pitenula}), and on the direction $(aB)_i$, (\ref{eq:pipit1}), to construct the projector orthogonal on them
\begin{equation}\label{eq:pite}
(\Pi_T)_i{}^j=\delta_i{}^j-(\Pi_0)_i{}^j-(\Pi_1)_i{}^j\, .
\end{equation}

For $a^2=0$, we have
\begin{equation}\label{eq:pete}
(P_T)_i{}^j=(\Pi_T)_i{}^j \Big|_{a^2=0}=\delta_i{}^j-(\Pi_0)_i{}^j-(P_1)_i{}^j\, .
\end{equation}
For $\tilde a^2=0$, $\Pi_0$ and $\Pi_T$ are singular, but it is useful to introduce projector
\begin{equation}\label{eq:pete1}
(\hat P_T)_i{}^j=\delta_i{}^j-(P_0)_i{}^j-(\hat P_1)_i{}^j\, .
\end{equation}

In the case when $a^2=\frac{1}{\alpha}$ we have
\begin{equation}\label{eq:starpt1}
({}^\star P_1)_i{}^j=(\Pi_1)_i{}^j\Big|_{a^2=\frac{1}{\alpha}}=\frac{4\alpha}{\alpha \tilde a^2-1}(Ba)_i(\tilde aB)^j\, ,
\end{equation}
\begin{equation}\label{eq:starpt}
({}^\star P_T)_i{}^j=(\Pi_T)_i{}^j\Big|_{a^2=\frac{1}{\alpha}}=\delta_i{}^j-({}^\star P_0)_i{}^j-({}^\star P_1)_i{}^j\, ,
\end{equation}
and similarly for $\tilde a^2=\frac{1}{\alpha}$ we get
\begin{equation}
({}^\star \hat P_0)_i{}^j=(\Pi_0)_i{}^j\Big|_{\tilde a^2=\frac{1}{\alpha}}=\alpha a_i \tilde a^j\, ,\quad ({}^\star \hat P_1)_i{}^j=(\Pi_1)_i{}^j\Big|_{\tilde a^2=\frac{1}{\alpha}}=\frac{4\alpha}{1-\alpha a^2}(Ba)_i(\tilde aB)^j\, ,
\end{equation}
\begin{equation}\label{eq:starptkapa}
\quad ({}^\star \hat P_T)_i{}^j=(\Pi_T)_i{}^j\Big|_{\tilde a^2=\frac{1}{\alpha}}=\delta_i{}^j-({}^\star \hat P_0)_i{}^j-({}^\star \hat P_1)_i{}^j\, .
\end{equation}

It is useful to introduce following notation for the projections of vectors $x^i$ and $\pi_i$
\begin{equation}\label{eq:notacija}
x_0=(n_0)_i x^i=a_i x^i\, ,\quad x_1=(n_1)_i x^i=(aB)_i x^i\, ,\quad \pi_0=\tilde n_0^i \pi_i=\tilde a^i \pi_i\, ,\quad \pi_1=\tilde n_1^i \pi_i=(\tilde aB)^i \pi_i\, .
\end{equation}

\end{document}